\documentclass[aps,prl,reprint,superscriptaddress,nobibnotes]{revtex4-2}
\usepackage{graphicx}
\usepackage{amsmath}
\usepackage{amssymb}
\usepackage{epstopdf}
\usepackage{color}
\usepackage{dsfont}
\usepackage{graphicx,wrapfig}
\usepackage{sidecap}
\usepackage{caption}
\usepackage{graphicx,amsfonts,amsmath,bm,bbm,amssymb,amsthm,color,url}
\DeclareMathOperator{\sinc}{sinc}

\usepackage[pdftex,bookmarks,colorlinks]{hyperref}
\usepackage{orcidlink}
\usepackage{braket}

\newcommand{\dd}{\mathrm{\,d}}

\begin{document}
\title{Teleportation scheme for the complete state of light at the example of coherent states}
\author{Tanita Permaul \orcidlink{0000-0003-2614-6182}}
\affiliation{School of Chemistry and Physics, University of KwaZulu-Natal, Durban 4000, South Africa}
\author{Arijit Dutta \orcidlink{0000-0001-8517-0073}}
\affiliation{School of Chemistry and Physics, University of KwaZulu-Natal, Durban 4000, South Africa}
\affiliation{Centre for Quantum Engineering, Research and Education, TCG CREST, Kolkata 700091, India}
\author{Filippus S. Roux \orcidlink{0000-0001-9624-4189}}
\affiliation{School of Chemistry and Physics, University of KwaZulu-Natal, Durban 4000, South Africa}
\author{Thomas Konrad \orcidlink{0000-0001-9380-7441}}
\email{konradt@ukzn.ac.za}
\affiliation{School of Chemistry and Physics, University of KwaZulu-Natal, Durban 4000, South Africa}
\affiliation{National Institute of Theoretical and Computational Sciences (NITheCS), KwaZulu-Natal, South Africa}

\begin{abstract}  
	We present a scheme to teleport both the spatial and number-of-photons degrees of freedom of light. This is achieved by teleportation of each pixel of a target image. We take coherent states as the input and demonstrate the scheme for the cases with ideal and realistic entanglement resources.
\end{abstract}

\maketitle

Quantum teleportation refers to the process which transfers the unknown state of a quantum system to another remote system. This process was first described for discrete variables,  in terms of the transfer of the states of two-level systems (qubits)\cite{bennettel}. The first experimental demonstration teleported polarisation qubits \cite{bouwmeester1997experimental}. The teleportation of multiple degrees of freedom of a single photon involved polarisation and 2-dimensional orbital angular momentum information \cite{wang2015quantum}. Only recently has higher-dimensional teleportation with spatial modes of light been experimentally realised \cite{sephton2023quantum, qiu2023remote}.

The teleportation of the number-of-photons degree of freedom was formulated in terms of continuous variables (quadrature amplitudes) for a single mode of light \cite{vaidman1994, Braunstein0}. An advantage of this method compared to schemes for discrete variables is that the teleportation process can be implemented with linear optical elements deterministically \cite{lutkenhaus1999} and efficiently \cite{goyal2014qudit}. However, only a single mode of the electromagnetic field can be transferred. A scheme was proposed to teleport multimode electromagnetic fields using continuous variable teleportation \cite{sokolov2001}. Nevertheless, no teleportation scheme so far has been designed to teleport the complete quantum state of light.  

Here we present a scheme to teleport a coherent state of light of unknown amplitude and spatial mode.  Such a teleportation, if successful,  can make any holographic (amplitude and phase) image appear at a distant plane without the information about the image traversing the space between sender and receiver.  Moreover, it turns out that the same method also remotely generates optical images that are carried by an arbitrary superposition of photon number (Fock) states. In addition, a sequence of such teleportations would allow an observer to monitor an object from afar, or to receive a holographic movie based on quantum teleportation. 
\begin{figure}
	\includegraphics[width=0.9\linewidth]{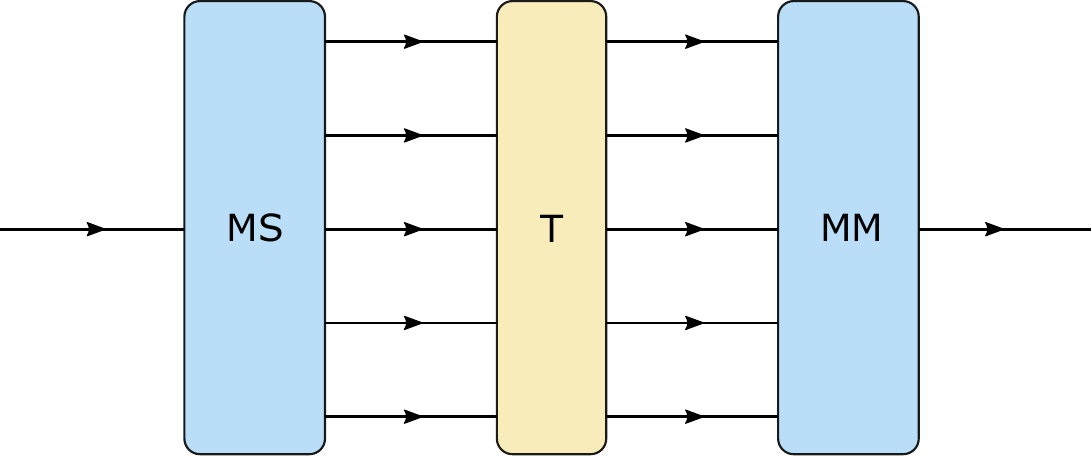}
	\caption{\label{fig:MS1}Schematic to teleport the complete state of light by transferring each mode to a different path and individually teleporting it. This scheme needs as many teleportation devices as there are modes. }
\end{figure}
\begin{figure}
	\includegraphics[width=0.8\linewidth]{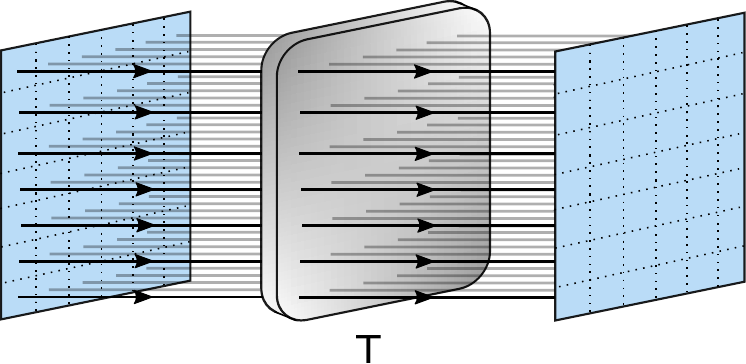}
	\caption{\label{fig:pixel}Transfer of a light mode from an input to an output screen by teleportation of the information carried by each pixel. A single source of squeezed states entangles input and output pixels.}
\end{figure}

The proposed teleportation scheme  is based on two properties of light.  Firstly, spatial light modes can be expressed in terms of a basis of distinct (orthogonal) modes. This implies that a coherent state of any spatial mode can be decomposed (e.g.\ by means of a mode sorter) into a tensor product of coherent states of the individual basis modes. A mode sorter is a device that maps each mode into a different optical path and thus spatially separates them. Each of these coherent states can then be teleported separately and the components can be synthesised by the receiver, using a mode sorter in reverse, to obtain the original coherent state, as illustrated in Fig.~\ref{fig:MS1}. Secondly, by the same principle, certain entangled (squeezed) states can be decomposed into a tensor product of individual two-mode entangled (squeezed) states, each of which can serve as the entanglement resource to teleport a corresponding basis mode.  Applying both properties to a certain basis that describes light in terms of pixels enables efficient teleportation of an input electromagnetic field pixel by pixel as shown in Fig.~\ref{fig:pixel}.  	

The state of the electromagnetic field in each pixel is a coherent state according to the first property and can be teleported using continuous variable (CV) teleportation. For this purpose, Charlie's input light field is superposed with Alice's light field on a beam splitter and both continuous variables (canonically conjugate variables) are measured for each pixel to displace Bob's light field accordingly. The detection can be efficiently carried out with CCD arrays, while the displacement field on Bob's side can be produced by a spatial light modulator. This process is described in detail later.

We start by demonstrating the first property. Charlie provides a coherent state of an arbitrary spatial mode $\mathcal{C}$ with creation operator $\hat{a}^\dag_{\mathcal{C}}$. The spatial mode is represented on a transversal plane at propagation distance z=0 by a normalised mode function $c (\textbf{x})$ that assigns an amplitude and phase of the electric field to each transversal position $\textbf{x}\in\mathbb{R}^2$. The corresponding angular spectrum (obtained by a Fourier transform) can be expressed in terms of basis modes by $C({\bf{k}}) = \sum_j \gamma_j F_{j} ({\bf{k}})$, where $\textbf{k}=(k_x, k_y)$ is the transverse part of the wave vector and $\gamma_j$ are complex valued coefficients. Then, the  creation operator  $\hat{a}^\dag_{\mathcal{C}}$ can be decomposed into a  sum of creation operators of the basis modes, 
\begin{eqnarray}
	\hat{a}^\dag_{\mathcal{C}} = \sum_{j} \gamma_j \hat{a}^\dag_{j}\quad\mbox{with}\,\, [\hat{a}_{j}, \hat{a}^\dag_{k}]=\delta_{jk}\label{adaggerC},
\end{eqnarray}
where $\hat{a}_j^\dagger = (1/(2\pi)^2) \int \hat{a}^{\dag}(\textbf{k}) F_j(\textbf{k})  \dd^2\textbf{k}$.
Since annihilation and creation operators of different basis modes commute, the initial coherent state $\ket{\alpha}_C$ of Charlie factorises into a tensor product of coherent states of individual basis modes,
\begin{eqnarray}\label{stateC}
	&\ket{\alpha}_C&=\hat{D}_C(\alpha)\ket{0}=\exp(\alpha \hat{a}^\dag_{\mathcal{C}}-\alpha^* \hat{a}_{\mathcal{C}} )\ket{0} \nonumber \\
	&&=\exp\left( \alpha \sum_j \gamma_j \hat{a}^{\dag}_{j} - \text{h.c.} \right) \ket{0}\\
	&&= \prod_j \exp\left( \alpha \gamma_j \hat{a}^{\dag}_{j} - \text{h.c.}\right) \ket{0} \nonumber =\bigotimes_j \ket{\alpha \gamma_j}_j,
\end{eqnarray}
where $\hat{D} (\alpha)$ is a displacement operator and h.c. refers to the Hermitian conjugate. Physically, the resulting  product of states means that any coherent state  of a light mode can be decomposed into a multitude of coexisting coherent states of basis modes. This can be tested using a mode sorter combined with homodyne detections to measure the complex amplitude of the coherent state for each basis mode. Moreover, any coherent state $\ket{\alpha}_C$ can also be synthesised by superposing the coherent states in the basis modes by means of a suitable multiport. For example, inserting coherent state $\ket{\alpha}_{1}$ in the first basis mode and coherent state $\ket{\beta}_{2}$ carrying the second basis mode into the input ports of a balanced beamsplitter results in the product state $\ket{\alpha/\sqrt{2}}_{1}\ket{\beta/\sqrt{2}}_{2}$ at one of its output ports.

The same factorisation is obtained in the pixel mode basis as explained below. Any light field can be decomposed with respect to the partition of a transversal plane into disjoint small areas (pixels) $A_j$ of equal size $A$, i.e.\ $\bigcup_i A_i = \mathbb{R}^2$ with $A_i\cap A_j = \delta_{ij}A_i$.  In position representation the $j$-th pixel mode is a normalised binary  function $f_j({\bf x})$ which assumes the value $1/A^{1/2}$ for positions ${\bf x}\in A_j$ in the pixel and $0$ outside. This leads to a decomposition of the mode $c(\textbf{x})$ into pixel modes based on the identity, $c(\textbf{x})=A^{1/2}\sum_j c(\textbf{x}) f_j(\textbf{x}) $. A faithful pixel representation assumes that the pixels are sufficiently small so that the mode value $c(\textbf{x})$ is constant over each pixel area $A_j$, i.e.\ $c(\textbf{x})= c_j$, and thus $c(\textbf{x})=\sum_j \gamma_j f_j(\textbf{x}) $ with $\gamma_j= c_j A^{1/2}  $. The state of a light field  with a single photon (elementary excitation) of the $j$-th pixel mode, $\ket{1}_{j} = \hat{a}^\dagger_j \ket{0}$, is expressed by a creation operator $\hat{a}^\dagger_j$, which can be decomposed into creation operators $\hat{a}^\dagger (\textbf{x})$ for a photon from point sources in each point of the pixel $\textbf{x}\in A_j$:
\begin{equation}
	\begin{split}
		\hat{a}_{j}^\dagger &= \int \hat{a}^{\dag}(\textbf{x})  f_j(\textbf{x})  \dd^2\textbf{x},  \\
		&=\frac{1}{(2\pi)^2} \int \hat{a}^{\dag}(\textbf{k}) F_j(\textbf{k})  \dd^2\textbf{k}, \\
	\end{split}
\end{equation}
where we used the Fourier transform of the creation operator of a photon from point source $\textbf{x}$ in terms of creation operators of plane waves with transversal wave vector $\textbf{k}$,  	 \begin{equation}\hat{a}^{\dag}(\textbf{x})=  \frac{1}{(2\pi)^2} \int \hat{a}^{\dag}(\textbf{k})   \text{exp}[i(\textbf{x} \cdot\textbf{k})] \dd^2\textbf{k},	
\end{equation}
and defined the angular spectrum of the $j$th pixel mode 
\begin{equation}
	F_j(\textbf{k}) \equiv \int f_j(\textbf{x})  \text{exp}[i(\textbf{x} \cdot\textbf{k})]\dd^2\textbf{x}.
\end{equation}
With these tools the creation operator of Charlie's mode $C(\textbf{k})=\sum_j \gamma_j F_j(\textbf{k})$ can be represented in terms of creation operators  of the corresponding pixel modes as expressed in Eq.\ (\ref{adaggerC}). This implies the decomposition (\ref{stateC}) of Charlie's state as a product of coherent states of the pixel modes. 	
\begin{figure}[ht]
	\centerline{\includegraphics[width=\linewidth]{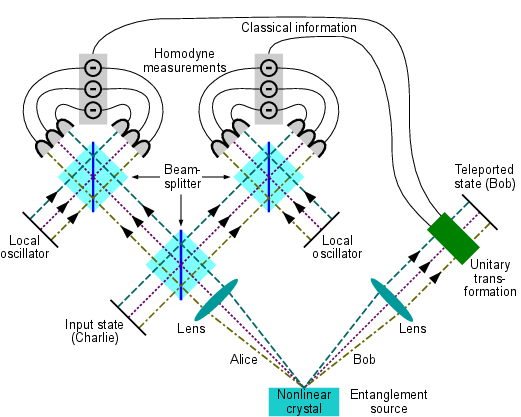}}
	\caption{\label{fig:ray} Ray diagram schematic to illustrate teleportation of multiple pixels at the same time. Here correlated pixels are identified by different coloured(dashed) lines.  }
\end{figure}
The second property requires the factorisation of the entangled state. In our context of CV teleportation, this translates to showing that the squeezed vacuum state can be expressed as a tensor product of two-mode squeezed states. An ideal two-mode squeezed vacuum state shared between Alice and Bob is of the form 
\begin{eqnarray}
	\hat{S}_{A,B}(\xi)\ket{0}=\mathrm{sech}r\ \sum_{n=0}^{\infty} (-e^{i\theta} \mathrm{tanh}r)^n  \ket{n}_A\ket{n}_B ,
\end{eqnarray}
where $\hat{S}_{A,B}(\xi)=\mathrm{exp}(\xi^* \hat{a} \hat{b}-\xi \hat{a}^{\dag}\hat{b}^{\dag} )$ is the twin-beam squeezing operator with squeezing parameter $\xi=r e^{i\theta}$. A more general description of the squeezing operator employs a squeezing kernel $h(\textbf{k}_1,\textbf{k}_2)$ instead of a squeezing parameter and reads 
\begin{equation}\label{biphotonkernel}
	\hat{S}_{A,B}= \text{exp}\left[\int (\hat{a}^\dagger(\textbf{k}_1) h(\textbf{k}_1,\textbf{k}_2) \hat{b}^\dagger(\textbf{k}_2) - \textrm{h.c.} )\frac{\dd^2\textbf{k}_1 \dd^2\textbf{k}_2 }{(2\pi)^4}\right].
\end{equation}

For maximal entanglement between Alice and Bob, the squeezing kernel is given by $h(\textbf{k}_1,\textbf{k}_2)= \xi (2\pi)^2 \delta(\textbf{k}_1+\textbf{k}_2)$ with squeezing parameter $\xi$. We can express the delta kernel in terms of basis modes in Fourier representation $F_j(\textbf{k})$ by
\begin{equation}\label{delapprox}
	\begin{split}
		(2\pi)^2 \delta(\textbf{k}_1+\textbf{k}_2) & =   \sum_{j}  F_j^*(-\textbf{k}_1)  F_j (\textbf{k}_2).  \\
	\end{split}
\end{equation}
In Dirac notation, this follows from expressing the completeness relation of basis modes $\mathcal{F}_j$, $\mathbb{I}=\sum_j \ket{\mathcal{F}_j}\bra{\mathcal{F}_j}$, in the momentum (angular frequency) representation, i.e. $\braket{\textbf{k}|\textbf{k}'}=\sum_j \braket{\textbf{k}|\mathcal{F}_j}\braket{\mathcal{F}_j|\textbf{k}'} $ with $\braket{\textbf{k}|\mathcal{F}_j}\equiv F_j(\textbf{k})$ and $\braket{\mathcal{F}_j|\textbf{k}'}=F^*(\textbf{k}')=F(-\textbf{k}')$.
      
Hence, we obtain for part of the exponent of the squeezing operator \eqref{biphotonkernel}
\begin{eqnarray}
	&&\int \hat{a}^\dag(\textbf{k}_1) h(\textbf{k}_1,\textbf{k}_2) \hat{b}^\dag(\textbf{k}_2) \frac{\dd^2\textbf{k}_1 \dd^2\textbf{k}_2 }{(2\pi)^4} \nonumber \\
	&&= \xi \int \sum_j \hat{a}^\dag(\textbf{k}_1) F_{j}(\textbf{k}_1)  F_j (\textbf{k}_2) \hat{b}^\dag(\textbf{k}_2) \frac{\dd^2\textbf{k}_1 \dd^2\textbf{k}_2 }{(2\pi)^4} \nonumber \\
	&&= \xi \sum_j \hat{a}_{j}^\dagger  \hat{b}_{j}^\dagger.  \label{lasteqn}
\end{eqnarray}
Therefore the squeezed state factorises as follows:
\begin{eqnarray}\label{sqstate}
&&\hat{S}_{A,B}\ket{0}=\mathrm{exp}\left[\xi \sum_j\left(\hat{a}_{j} \hat{b}_{j}- \hat{a}^{\dag}_{j}\hat{b}^{\dag}_{j}\right) \right] \ket{0}\nonumber\\
&=& \prod_j \mathrm{exp}\left(\xi \hat{a}_{j} \hat{b}_{j}- \xi \hat{a}^{\dag}_{j}\hat{b}^{\dag}_{j}\right) \ket{0}\nonumber\\
&=& \bigotimes_j\    \mathrm{sech}\xi\ 
		\sum_{n_j=0}^{\infty} ( \mathrm{tanh}\xi)^{n_j}\ket{n_{j}}_{A_{j}}\ket{n_{j}}_{B_{j}}. 
\end{eqnarray}	
In the case of the pixel basis, the equality in Eq.\ (\ref{delapprox}) holds approximately in the regime where the pixel size is much smaller than the scale of the variations (wavelength)  in the far-field  of the light beams. Therefore Eq.~(\ref{sqstate}) shows that the squeezed vacuum state can be decomposed into a tensor product of two-mode squeezed vacuum states for each correlated pair of pixels. Here for convenience we denote the correlated pixel pairs of Alice and Bob with the same index $j$. We discuss these correlations in detail in the Supplementary Information (SI). 

The squeezing parameter $\xi$ is the same for all pixels due to the choice of the ideal delta-correlated squeezing kernel. In the general case with a non-ideal squeezing kernel, different pixels would have different squeezing parameters $\xi_j$. In addition, we assume below that $\xi_j$ is a real number $r_j$.

We explained how the teleportation channel can be expressed as a tensor product of the individual teleportation channels between corresponding pixels. Hence we now analyse the process to transfer information from a single pixel to another, which is identical to continuous variable teleportation with one spatial mode~\cite{Braunstein0, Milburn, Braunstein1, Hofmann, Braunstein2, Pirandola}. 

CV teleportation consists of a Bell measurement followed by a unitary transformation (feedback) acting on Bob's state. 
The initial state of Alice's, Bob's and Charlie's electromagnetic field on their $j$-th pixels can be expressed as,
\begin{equation}\label{fidelity14}
	\begin{split}
			&\ket{\Psi_{\text{init}}}_{ABC}=|\Phi\rangle_{AB}\otimes |\tilde{\alpha}\rangle_C,\\
	\end{split}
\end{equation}
where $\tilde{\alpha}\equiv\alpha\gamma_j$ (compare Eq.\ (\ref{stateC})) and 
\begin{equation}
	\ket{\Phi}_{AB}=\text{sech}(r)\sum_n\text{tanh}^n(r)|n\rangle_{A}|n\rangle_{B}
\end{equation}
is the $j$-th factor of the squeezed state in Eq.\ (\ref{sqstate}) with pixel index $j$ omitted.  
	 
The Bell measurement is implemented by means of homodyne detections and projects Alice's and Charlie's components of the initial state onto a generalised Bell state,  
\begin{equation}
	\label{fidelity16}
	\ket{\Psi(\beta)}_{AC}= \frac{1}{\sqrt{\pi}}\sum_s \hat{D}_{C}(\beta)|s\rangle_A|s\rangle_{C},
\end{equation}
where $\beta=\beta_{\text{R}}+i \beta_{\text{I}}$ incorporates the measurement results of the homodyne detections as detailed in the SI.
The transformation affects Bob's system due to its entanglement with Alice's system,
\begin{equation}
	\begin{split}
	\ket{\Psi_{\text{init}}}_{ABC}\xrightarrow{\text{result}\ ``\beta"}& \frac{1}{\sqrt{p(\beta)}}\  _{AC}\braket{\Psi(\beta)|\Psi_{\text{init}}}_{ABC} \\
	&= \ket{\zeta(\beta)}_B.
	\end{split}
\end{equation}
Here $p(\beta)$ is the probability to obtain measurement result $\beta$ and  Bob's state, modulo a global phase factor, is  a coherent state $\ket{\zeta(\beta)}$ with amplitude
\begin{equation}
	\zeta(\beta)=\text{tanh}(r)(\tilde{\alpha}-\beta).
\end{equation}
It resembles Charlie's input state $\ket{\tilde{\alpha}}$ with amplitude shifted by the measurement result $\beta$ and reduced by factor $ \text{tanh}(r) \in [-1,1]$.  
To compensate the shift, Bob applies an opposite displacement $\hat{D}(\beta)$ and obtains the coherent state $\ket{\zeta(\beta) +\beta}$ with amplitude 
\begin{equation}
	\zeta (\beta) +\beta=\text{tanh}(r)\tilde{\alpha}+(1-\text{tanh}(r))\beta .
\end{equation}
In the limit of infinite squeezing, $r
\rightarrow \infty \Rightarrow\tanh r =1$, the input state $\ket{\tilde{\alpha}}$ is teleported perfectly from Charlie to Bob. Without squeezing ($r=\tanh r = 0$)
no information is teleported. To measure how close the output state,  $\ket{\psi_{\text{out}}}\equiv|\zeta(\beta)+\beta \rangle$, is to Charlie's initial state $\ket{\psi_{\text{in}}}\equiv\ket{\tilde{\alpha}}$, we calculate the fidelity for one pixel,
\begin{equation}\label{fidelity190}
	\begin{split}
		F(\beta)&=|\langle \psi_{\text{out}}|\psi_{\text{in}} \rangle|^2\\
		&=\text{exp}\left(-[1-\text{tanh}(r)]^2|\tilde{\alpha}-\beta|^2 \right). \\
	\end{split}
\end{equation}
After averaging over all possible values of $\beta$, we obtain the average fidelity for one pixel,
\begin{equation}\label{fidelity21}
	\begin{split}
		&F=\int p(\beta)F(\beta) \dd^2\beta
		= \frac{1+\text{tanh}(r)}{2}.\\
	\end{split}
\end{equation}

Now we consider teleportation between corresponding pixels with spontaneous parametric down-conversion as the source of entanglement. A more detailed discussion of the following calculation is presented in the SI. 

We study down-conversion with a pump beam incident on a nonlinear crystal of length $L$ to produce entangled pairs of photons, known as signal and idler photons. The pump beam has a transverse Gaussian profile,
\begin{equation}
	\Phi(\textbf{k})=\sqrt{2\pi}w_{\text{p}} \varphi_0 \text{exp}\left(\frac{-|\textbf{k}|^2}{4}w_{\text{p}}^2\right),
\end{equation}
 where $w_{\text{p}}$ is the beam waist radius and $\varphi_0$ is a complex amplitude.
The squeezing kernel $h'(\textbf{k}_1,\textbf{k}_2)$ can be modelled by the pump beam spatial mode and the phase mismatch function,
\begin{equation}\label{kernelg}
	h'(\textbf{k}_1,\textbf{k}_2,z)=\Omega_0 \sqrt{\omega_1 \omega_2} \text{exp}\left( -\frac{w_{\text{p}}^2 |\textbf{k}_1+\textbf{k}_2|^2}{4}  + i \Delta k_z z \right) ,
\end{equation}
where $\Omega_0$ is a combination of dimensional parameters obtained under the monochromatic condition, $\omega_{1,2}$ denotes the down-converted frequencies, and $\Delta k_z$ is the $z$-component wavevector mismatch. This model represents a good approximation for a low number of down converted photons.
 
\begin{figure}[ht]
	\centerline{\includegraphics{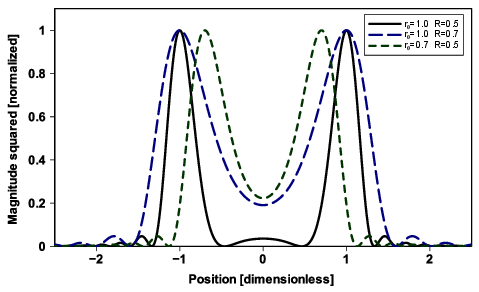}}
	\caption{The normalised values of $|\tilde{\eta}(\mathbf{x}_0)|^2$ along any direction in the output plane for the different combinations of ring radius $r_0$ and ring width $R$;  $r_0=1$ and $R=0.5$,  $r_0=1$ and $R=0.7$, and $r_0=0.7$ and $R=0.5$.}
	\label{f3k0}
\end{figure}

This kernel can be substituted into Eq.~(\ref{biphotonkernel}) to obtain the corresponding two-mode squeezed operator. Instead of using the spectra $F_j(\textbf{k})$ of flat pixel modes defined above, for computational ease we replace them with a Gaussian profile $M(\textbf{k}_i; \textbf{k}_0)$, centered at $\textbf{k}_0$ with width $w_0$, in order to integrate them with the non-ideal squeezing kernel,
where
\begin{equation}
	M(\textbf{k}_i; \textbf{k}_0)=\sqrt{2\pi}w_0\text{exp}\left( \frac{-|\textbf{k}_i-\textbf{k}_0|^2}{4}w_0^2\right).
\end{equation}
Here it is assumed that the mode size of the pixel modes in the detector plane is much smaller than that of the pump, i.e. $w_0\ll w_{\text{p}}$. 
After this replacement, the two-mode squeezing operator becomes
\begin{equation}
	\label{detmodsqueezed}
	\hat{S}(\eta)=\text{exp}\left(\int \hat{a}^{\dagger}_{M}(\textbf{k}_0)\eta(\textbf{k}_0)\hat{b}^{\dagger }_{M}(-\textbf{k}_0) \dd^2\textbf{k}_0 -h.c.\right),
\end{equation}
where the overlap $\eta(\textbf{k}_0)$ between the non-ideal squeezing kernel and the pixel  modes is calculated, as shown in the SI, to be
\begin{equation}\label{fk0}
	\eta(\textbf{k}_0)=\Xi \sinc\left(\frac{|\textbf{k}_0|^2L}{k_{\text{p}}}-\frac{1}{2} L \chi\right).
\end{equation}
Here $k_{\text{p}}$ is the pump beam wavenumber, $\Xi$ is the effective squeezing parameter produced by the experimental conditions and $\chi$ is a parameter obtained by the non-collinear phase matching conditions (see SI). 
In the far-field, $|\textbf{k}_0|$ is replaced by $ |\textbf{x}_0|  k_{\text{p}}/(2f)$, thus 
\begin{equation}
	\eta(\textbf{k}_0)\rightarrow \tilde{\eta}(\textbf{x}_0)=\Xi \sinc\left(\frac{|\mathbf{x}_0|^2-r_0^2}{R^2} \right)
\end{equation}
where $\textbf{x}_0$ is the transverse coordinate vector,
\begin{equation}
	R = \frac{2f}{\sqrt{L k_{\text{p}}}}, ~~~ \text{and} ~~~
	r_0 = f\tan(\theta_{\text{d}}) .
\end{equation}
Here $f$ is the focal length of the lens and $\theta_{\text{d}}$ is the down-conversion angle. 
Therefore, $\tilde{\eta}(\mathbf{x}_0)$ is a ring-shaped squeezing function with radius $r_0$ and width $R$. The shape of the cross-section for different parameters is shown in Figure~\ref{f3k0}.

\begin{figure}[ht]
	\centerline{\includegraphics{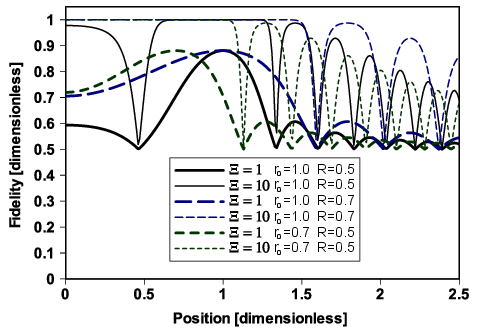}}
	\caption{The fidelity along an arbitrary direction, corresponding to the curves in Figure~\ref{f3k0} for $\Xi=1$ and $\Xi=10$.}
	\label{fidall}
\end{figure}

We compute the teleportation fidelity based on the expression in Eq.\ (\ref{fidelity21}) where $r$ is replaced by $|\tilde{\eta}(\mathbf{x}_0)|$ for the pixel at position $\textbf{x}_0$.
The resulting curves are plotted in Fig.\ \ref{fidall}, corresponding to those in Fig.\ \ref{f3k0} respectively, for squeezing parameters $\Xi=1$ and $\Xi=10$.

To calculate the total fidelity of the coherent state teleportation, we use the average pixel fidelity for $N$ pixels,
\begin{equation}
	F_{\text{image}}=\frac{1}{N} \sum_j F_j =\frac{1}{N}\sum_j \left(\frac{1+\text{tanh}(r_j)}{2} \right).
\end{equation}
The fidelity we calculate above gives an average value for how well the coherent amplitudes for each pixel and thus the amplitude and phase values of the input mode are teleported.
 		
We have shown that it is possible to teleport, with the scheme described above, an unknown spatial mode and complex amplitude of a coherent state efficiently i.e., with scalable resources. Our argument was based on that both -- the input state of light to be teleported and the entangled resource state -- can be decomposed into tensor products in terms of basis modes. Subject to these conditions Charlie's and Alice's basis modes can be spatially separated and teleported individually. We discussed the validity of the assumption based on a model of squeezed (entangled) light beams generated by parametric down conversion. 
Our scheme made use of pixel modes that are naturally spatially separated. They form in good approximation a basis for the incoming light field, if it varies on a spatial scale greater than the dimension of the pixels. Although we demonstrated teleportation of coherent states, we conjecture that the teleportation scheme presented here works for arbitrary superpositions of Fock states as will be discussed elsewhere. 
\linebreak

\textit{Acknowledgments}---
T.K. and A.D. are grateful for financial support by the South African Quantum Technology Initiative and the National Institute of Theoretical and Computational Sciences. T.P. acknowledges a PhD bursary by the National Research Foundation of South Africa. 
			
\appendix

\section{Appendix}

\section{Pixel correlations}
As shown in Fig.\ \ref{fig:ray} of the main text, the rays indicating correlated pixels on Alice and Bob's transverse planes are anti-correlated. We now explain why this physically occurs. Following Eq.\ (\ref{delapprox}) of the paper,
\begin{equation}
	\begin{split}
		(2\pi)^2 \delta(\textbf{k}_1+\textbf{k}_2) & =   \sum_{j}  F_j^*(-\textbf{k}_1)  F_j (\textbf{k}_2),  \\
	\end{split}
\end{equation}
we use a mathematical simplification $F_j^*(\textbf{k})=F_j(-\textbf{k})$, so we can we can express the delta-correlated kernel as $(2\pi)^2 \delta(\textbf{k}_1+\textbf{k}_2) =   \sum_{j}  F_j(\textbf{k}_1)  F_j (\textbf{k}_2)$. Here we see that Alice and Bob's pixel modes share the same pixel ordering, which would be true if we consider Alice and Bob's fields on the same transverse plane.  However, since we know that the transverse momentum of the down-converted fields are anticorrelated, the transverse planes of Alice and Bob are complex conjugates of each other. As the creation operator for Alice's field in Eqn.\ (\ref{sqstate}) of the paper is in terms of postive $\textbf{k}_1$, we must reverse the pixel index for Alice's transverse field and therefore 
\begin{equation}
	(2\pi)^2 \delta(\textbf{k}_1+\textbf{k}_2) =   \sum_{j=1}^N  F_{N-j}^*(\textbf{k}_1)  F_j (\textbf{k}_2),
\end{equation}
where $N$ is the total number of pixel modes. We show the reversed pixel indexing in Fig.\ \ref{reversegrid} below. Therefore the $N-j$-th and $j$-th pixels of Alice and Bob are correlated. 
\begin{figure}[ht]
	\centerline{\includegraphics[width=0.7\linewidth]{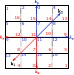}}
	\caption{The transverse planes of Alice (red) and Bob (blue) for their respective wave vectors $\textbf{k}_1$ and $\textbf{k}_2$ for $N=16$ pixels. The pixel numbering for Bob's plane is shown in the top left corner of each pixel while for Alice's plane it is in the bottom right. We therefore see how the $N-j$ and $j$ pixels are correlated if Alice has a reversed pixel ordering.   }
	\label{reversegrid}
\end{figure}

\section{CV teleportation}
\begin{figure}[t]
	\includegraphics[scale=0.5]{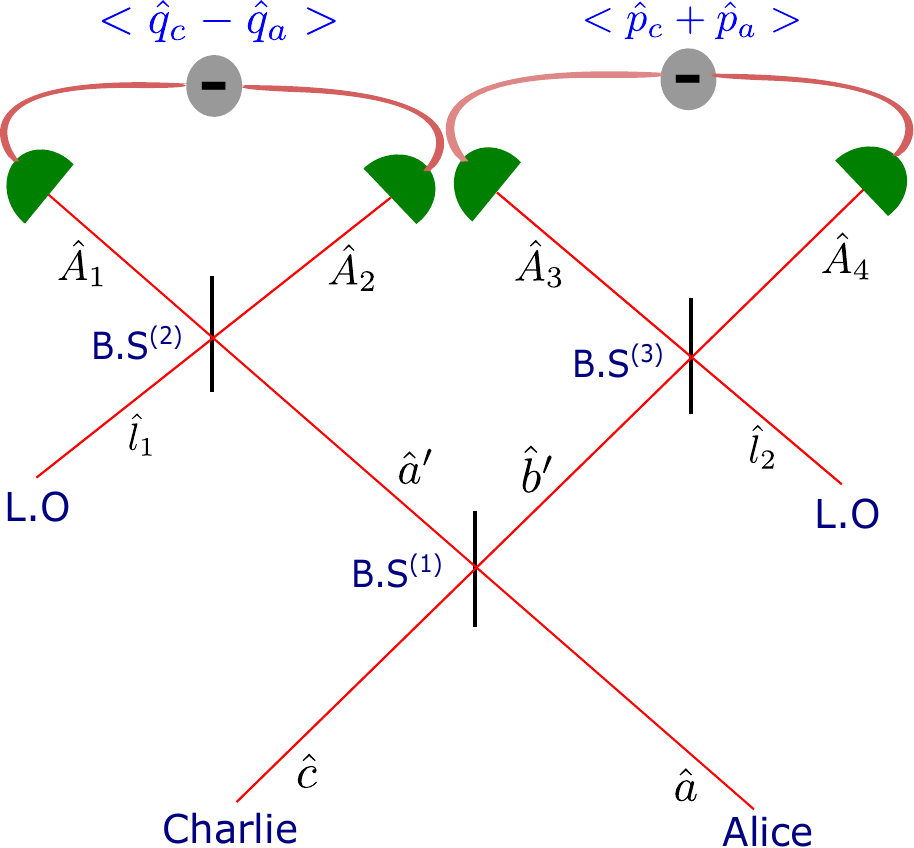}
	\centering
	\caption{homodyne detections in the teleportation scheme}
	\label{real}
\end{figure}
In the quantum teleportation scheme, Charlie wants to send an unknown state from her side to Bob's side by using entanglement as a resource. Alice and Bob share pairs of two-mode squeezed vacuum states. For such a two-mode squeezed vacuum state the annihilation operator associated with Alice's mode is denoted by $\hat{a}_A,$ and the corresponding annihilation operator for Bob's mode is $\hat{a}_B$ (see Fig.~\ref{real}). 
$\hat{a}_C$  is the annihilation operator associated with Charlie's mode. Two input modes with the annihilation operators $\hat{a}_A,$ and $\hat{a}_C$  are combined at $\text{B.S}^{(1)}.$ The two output modes are represented by $\hat{b}_A,$ and $\hat{b}_C.$ The transformation relations read
\begin{equation}
	\begin{split}
		\label{homodyne00}
		\hat{b}_A=\frac{\hat{a}_C-\hat{a}_A}{\sqrt{2}}, \hat{b}_C=\frac{\hat{a}_A+\hat{a}_C}{\sqrt{2}}.\\
	\end{split}
\end{equation}

To exploit the entanglement for reconstructing the unknown state on Bob's side, Alice needs to perform two homodyne detection measurements with the proper choice of local oscillator states. Alice estimates $\langle\frac{\hat{q}_C-\hat{q}_A}{\sqrt{2}}\rangle$ by measuring the photon number difference at the two outputs of $\text{B.S}^{(2)}.$  Here the annihilation operator corresponds to the local-oscillator mode is represented by $\hat{l}_1.$ Similarly, $\langle\frac{\hat{p}_A+\hat{p}_C}{\sqrt{2}}\rangle$ is obtained by measuring photon number difference at the two outputs of $\text{B.S}^{(3)}.$ For the homodyne detection at $\text{B.S}^{(3)},$ the annihilation operator associated with the local-oscillator mode is  $\hat{l}_2.$ The annihilation operators for the two output modes of $\text{B.S}^{(2)}$ are written as follows 
\begin{equation}
	\begin{split}
		\label{homodyne000}
		\hat{A}_{1}&=\frac{1}{\sqrt{2}}(\hat{l}_1-\frac{\hat{a}_C-\hat{a}_A}{\sqrt{2}}),\\
		\hat{A}_{2}&=\frac{1}{\sqrt{2}}(\hat{l}_1+\frac{\hat{a}_C-\hat{a}_A}{\sqrt{2}}).\\
	\end{split}
\end{equation}

From $\eqref{homodyne000},$ we obtain
\begin{equation}
	\begin{split}
		\label{homodyne1}
		\hat{A}_2^{\dagger}\hat{A}_2-\hat{A}_1^{\dagger}\hat{A}_1&=\frac{1}{\sqrt{2}}((\hat{l}_1^{\dagger}\hat{a}_C+\hat{l}_1\hat{a}_C^{\dagger})-(\hat{l}_1^{\dagger}\hat{a}_A+\hat{l}_1\hat{a}_A^{\dagger})).\\
	\end{split}
\end{equation}
Similarly, the difference in photon number operators at the two output modes of $\text{B.S}^{(3)}$ is written as
\begin{equation}
	\begin{split}
		\label{homodyne2}
		\hat{A}_{4}^{\dagger}\hat{A}_{4}-\hat{A}_{3}^{\dagger}\hat{A}_{3}&=\frac{1}{\sqrt{2}}((\hat{l}_2^{\dagger}\hat{a}_C+\hat{l}_2\hat{a}_C^{\dagger})+(\hat{l}_2^{\dagger}\hat{a}_A+\hat{l}_2\hat{a}_A^{\dagger})).\\
	\end{split}
\end{equation}
One can estimate photocurrents by measuring the difference in average photon numbers at the two output modes of $\text{B.S}^{(2)},$ and $\text{B.S}^{(3)}.$ Also, we define quadrature operators as
\begin{eqnarray}
	\hat{q}_A=\frac{\hat{a}_A^{\dagger}+\hat{a}_A}{\sqrt{2}},\ \ \hat{p}_A=i\frac{\hat{a}_A^{\dagger}-\hat{a}_A}{\sqrt{2}},
\end{eqnarray}
and so on. If we consider local-oscillator modes are in coherent states $|\alpha_j\rangle,$ where $j\in\{1,2\},$ then the expectation values of $\hat{l}_j,$ and $\hat{l}_j^{\dagger}$ on  $|\alpha_j\rangle'$s with $\alpha_j=|\alpha_j|e^{i\phi_j}$ are given as follows: $\langle\alpha_j|\hat{l}_j|\alpha_j\rangle=\alpha_j$ and $\langle\alpha_j|\hat{l}_j^{\dagger}|\alpha_j\rangle=\alpha_j^*,$ where $\alpha^*_j$ is complex conjugate of $\alpha_j.$
Thus, one obtains,
\begin{eqnarray}
	\label{homodyne10}
	i_{q_-}=&\langle\hat{A}_2^{\dagger}\hat{A}_2-\hat{A}_1^{\dagger}\hat{A}_1\rangle_{\phi_1=0}=|\alpha_1|\langle\hat{q}_C-\hat{q}_A\rangle,\\
\end{eqnarray}
and
\begin{equation}
	\begin{split}
		\label{homodyne100}
		i_{p_+}=&\langle\hat{A}_4^{\dagger}\hat{A}_4-\hat{A}_3^{\dagger}\hat{A}_3\rangle_{\phi_2=\pi/2}=|\alpha_2|\langle\hat{p}_A+\hat{p}_C\rangle,\\
	\end{split}
\end{equation}
where $i_{q_-}$ and $i_{p_+}$ are the photocurrents. 
Mathematically, the simultaneous homodyne detection measurements on Alice's side project the combined state of the two modes on a common eigenstate corresponding to the observables $\frac{\hat{q}_C-\hat{q}_A}{\sqrt{2}},$ and $\frac{\hat{p}_A+\hat{p}_C}{\sqrt{2}}.$ After rescaling, one can represent the eigenstate as
\begin{equation}
	\begin{split}
		\label{eigenstate}
		\left|\frac{q'-q}{\sqrt{2}}, \frac{p+p'}{\sqrt{2}}\right\rangle&=\frac{1}{\sqrt{\pi}}\sum_{n=0}^{\infty}\hat{D}_C(\beta)|n\rangle_A|n\rangle_C,\\
	\end{split}
\end{equation}
where $\beta=\beta_R+i\beta_I=\frac{1}{\sqrt{2}}((q'-q)+i (p+p')).$ The formal proof is given below. 

We start our analysis by considering the action of an annihilation operator $\hat{a}_C$ on the state $\frac{1}{\sqrt{\pi}}\sum_{n=0}^{\infty}\hat{D}_C(\beta)|n\rangle_A|n\rangle_C.$ Precisely,
\begin{equation}
	\begin{split}
		\label{eigenstate1}
		&\hat{a}_C\frac{1}{\sqrt{\pi}}\sum_{n=0}^{\infty}\hat{D}_C(\beta)|n\rangle_A|n\rangle_C\\
		&=\frac{1}{\sqrt{\pi}}\sum_{n=0}^{\infty}\hat{a}_Ce^{-1/2|\beta|^2}e^{\beta\hat{a}_C^{\dagger}}e^{-\beta^*\hat{a}_C}\frac{\hat{a}_A^{\dagger n}\hat{a}_C^{\dagger n}}{n!}|0\rangle_A|0\rangle_C\\
		&=\frac{1}{\sqrt{\pi}}\sum_{n=0}^{\infty}e^{-1/2|\beta|^2}\hat{a}_Ce^{\beta\hat{c}^{\dagger}}e^{-\beta^*\hat{a}_C}\frac{\hat{a}_A^{\dagger n}\hat{a}_C^{\dagger n}}{n!}|0\rangle_A|0\rangle_C.\\
	\end{split}
\end{equation}
Now using the commutation relation  $[\hat{A},e^{\lambda\hat{B}}]=\lambda\hat{C}e^{\lambda\hat{B}},$ where $[\hat{A},\hat{B}]=\hat{C},$ we can replace $\hat{a}_Ce^{\beta\hat{a}_C^{\dagger}}$ with $\left(\beta e^{\beta\hat{a}_C^{\dagger}}+e^{\beta\hat{a}_C^{\dagger}}\hat{a}_C\right ).$ Inserting this in $\eqref{eigenstate1},$ we obtain
\begin{equation}
	\begin{split}
		\label{eigenstate2}
		&\hat{a}_C\frac{1}{\sqrt{\pi}}\sum_{n=0}^{\infty}\hat{D}_C(\beta)|n\rangle_A|n\rangle_C\\
		&= \frac{1}{\sqrt{\pi}}\sum_{n=0}^{\infty}e^{-1/2|\beta|^2}(\beta e^{\beta\hat{a}_C^{\dagger}}+e^{\beta\hat{a}_C^{\dagger}}\hat{a}_C) \\
		&\ \ \times e^{-\beta^*\hat{a}_C}\frac{\hat{a}_A^{\dagger n}\hat{a}_C^{\dagger n}}{n!}|0\rangle_A|0\rangle_{C}.\\
	\end{split}
\end{equation}
In the next step, we use the following commutation relation
\begin{equation}
	\begin{split}
		\label{eigenstate3}
		[\hat{a}_C,\hat{a}_C^{\dagger n}]&=n\hat{a}_C^{\dagger n-1}.\\
	\end{split}
\end{equation}
Replacing $\hat{a}_C\hat{a}_C^{\dagger n}$ with $(n\hat{a}_C^{\dagger n-1}+\hat{a}_C^{\dagger n}\hat{a}_C),$ on the right-hand side of the Eq.~$\eqref{eigenstate2},$ we get
\begin{equation}
	\begin{split}
		\label{eigenstate4}
		&\beta\frac{1}{\sqrt{\pi}}\sum_{n=0}^{\infty}\hat{D}_C(\beta)|n\rangle_A|n\rangle_C\\
		&+\hat{a}^{\dagger}\frac{1}{\sqrt{\pi}}\sum_{n=0}^{\infty}e^{-1/2|\beta|^2} e^{\beta\hat{a}_C^{\dagger}}
		e^{-\beta^*\hat{a}_C}\frac{\hat{a}_A^{\dagger n-1}\hat{a}_C^{\dagger n-1}}{n-1!}|0\rangle_A|0\rangle_{C}\\
		&=(\beta+\hat{a}_A^{\dagger})\frac{1}{\sqrt{\pi}}\sum_{n=0}^{\infty}\hat{D}_C(\beta)|n\rangle_A|n\rangle_C.\\
	\end{split}
\end{equation}
After inserting $\eqref{eigenstate4}$ on the right-hand side of $\eqref{eigenstate2}$ and rearranging, we get the following eigenvalue equation of the operator $\hat{a}_C-\hat{a}_A^{\dagger},$ 
\begin{equation}
	\begin{split}
		\label{eigenstate5}
		&(\hat{a}_C-\hat{a}_A^{\dagger})\frac{1}{\sqrt{\pi}}\sum_{n=0}^{\infty}\hat{D}_C(\beta)|n\rangle_A|n\rangle_C\\
		&=\beta\frac{1}{\sqrt{\pi}}\sum_{n=0}^{\infty}\hat{D}_C(\beta)|n\rangle_A|n\rangle_C,
	\end{split}
\end{equation}
where $\beta$ is the eigenvalue associated with the eigenstate $\frac{1}{\sqrt{\pi}}\sum_{n=0}^{\infty}\hat{D}_C(\beta)|n\rangle_A|n\rangle_C.$

Similarly one obtains, 
\begin{equation}
	\begin{split}
		\label{eigenstate6}
		&(\hat{a}_A-\hat{a}_C^{\dagger})\frac{1}{\sqrt{\pi}}\sum_{n=0}^{\infty}\hat{D}_C(\beta)|n\rangle_A|n\rangle_C\\
		&=-\beta^*\frac{1}{\sqrt{\pi}}\sum_{n=0}^{\infty}\hat{D}_C(\beta)|n\rangle_A|n\rangle_C.
	\end{split}
\end{equation}
Subtracting $\eqref{eigenstate5}$ from $\eqref{eigenstate6}$ we get
\begin{equation}
	\begin{split}
		\label{eigenstate7}
		&\frac{\hat{q}_C-\hat{q}_A}{\sqrt{2}}\frac{1}{\sqrt{\pi}}\sum_{n=0}^{\infty}\hat{D}_C(\beta)|n\rangle_A|n\rangle_C\\
		&=\beta_R\frac{1}{\sqrt{\pi}}\sum_{n=0}^{\infty}\hat{D}_C(\beta)|n\rangle_A|n\rangle_C,
	\end{split}
\end{equation}
where $\hat{q}_A=\frac{\hat{a}_A^{\dagger}+\hat{a}_A}{\sqrt{2}},$ and so on.
Similarly, adding $\eqref{eigenstate5}$ to $\eqref{eigenstate6}$ we obtain
\begin{equation}
	\begin{split}
		\label{eigenstate8}
		&\frac{\hat{p}_A+\hat{p}_C}{\sqrt{2}}\frac{1}{\sqrt{\pi}}\sum_{n=0}^{\infty}\hat{D}_C(\beta)|n\rangle_A|n\rangle_C\\
		&=\beta_I\frac{1}{\sqrt{\pi}}\sum_{n=0}^{\infty}\hat{D}_C(\beta)|n\rangle_A|n\rangle_C,
	\end{split}
\end{equation}
where $\hat{p}_A=i\frac{\hat{a}_A^{\dagger}-\hat{a}_A}{\sqrt{2}},$ and so on.

\section{Realistic entanglement source}	

We discuss in detail the calculation of the squeezing parameter obtained from spontaneous parametric down-conversion with a Gaussian pump. From Eq.\ (\ref{kernelg}), the down-conversion kernel is given by
\begin{eqnarray}
	&h'(\textbf{k}_1,\textbf{k}_2,z)=\Omega_1  \text{exp}\left( -\frac{w_{\text{p}}^2|\textbf{k}_1+\textbf{k}_2|^2}{4}  + i \Delta k_z \right) ,&
\end{eqnarray}
where $\Omega_1=\Omega_0\sqrt{\omega_1 \omega_2}$ and $w_{\text{p}}$ is the pump beam waist radius. Here $\omega$ denotes the frequencies of either the pump beam or the down-converted beams, $\Delta k_z $ is the phase mismatch condition in the $z$-component. This constant is proportional to the complex amplitude of the pump beam $\varphi_0$ and the nonlinear coefficient for Type-I phase matching of the nonlinear crystal $\sigma_{\text{ooe}}$. Note that all three beams are considered paraxial, but not necessarily collinear. For example, if the pump beam is assumed to propagate along the z-direction, the other two beams are not forced to propagate along the same direction. Thus, if the down conversion angle $\theta_{\text{d}}$ between the pump beam and signal/idler beam is not sufficiently small, then due to the non-collinearity of the beams,  
$\Delta k_z$ becomes
\begin{equation}
	\label{mismatch1}
	\Delta k_z=\frac{|\textbf{k}_1-\textbf{k}_2|^2}{2k_{\text{p}}}-\chi,
\end{equation}
where $|\textbf{k}_i-\textbf{k}_j|^2=(k_{ix}-k_{jx})^2+(k_{iy}-k_{jy})^2$,  $k_{\text{p}}$ is the pump wavenumber and
\begin{equation}
	\chi=\frac{k_{\text{d}}\sin^2\theta_{\text{d}}}{\cos\theta_{\text{d}}},
\end{equation}
with $k_{\text{d}}$ being the wavenumber associated to signal/idler beam in the degenerate case.
We now overlap the down-conversion kernel $h'(\textbf{k}_1,\textbf{k}_2,z)$ with detector modes that have a Gaussian profile with width $w_0$,
\begin{eqnarray}
	M(\textbf{k}_i,\textbf{k}_j)=\sqrt{2\pi}\ w_0^2 \text{exp}\left( -\frac{1}{4} w_0^2 |\textbf{k}_i -\textbf{k}_j|^2 \right).
\end{eqnarray}
We also consider monochromatic conditions so that we can ignored any temporal frequency dependences. After the integrals over the transverse wavevectors for the overlaps by the detector mode have been evaluated, but prior to the evaluation of the $z$-integration, we have
\begin{align}\label{eval0}
	&\eta(\mathbf{k}_a,\mathbf{k}_b) =  \int_{-L/2}^{L/2} \int M(\mathbf{k}_1;\mathbf{k}_a) h'(\mathbf{k}_1,\mathbf{k}_2,z) \nonumber\\
	& \hspace{2cm} \times M(\mathbf{k}_2;\mathbf{k}_b)\ \frac{\text{d}^2 k_1 \text{d}^2 k_2}{(2\pi)^4}\ \text{d}z \nonumber \\
	&= \int_{-L/2}^{L/2} \frac{2\Xi w_{\text{p}}^2}{(w_0^2+2w_{\text{p}}^2)L}\nonumber \\
	&\times
	\exp\left(-\frac{w_{\text{p}}^2 w_0^2 |\mathbf{k}_a+\mathbf{k}_b|^2}{4(w_0^2+2w_{\text{p}}^2)}  -\frac{i z |\mathbf{k}_a-\mathbf{k}_b|^2}{2k_{\text{p}}} +i z\chi \right)\ \text{d}z,
\end{align}
where we initially use different transverse displacement wavevectors $\mathbf{k}_a$ and $\mathbf{k}_b$ for the two modes, and express the prefactor and the exponent in leading order in $z$. The dimensionless parameter $\Xi$ is a combination of dimensional parameters obtained under the monochromatic condition, and proportional to the magnitude of the pump beam amplitude $\Xi \propto  |\varphi_0|$.

At this point, we'll assume that $w_{\text{p}}\gg w_0$, because the pixels are assumed to smaller than any feature size in the parameter functions of the initial state. Hence,
\begin{align}\label{eval1}
	\eta(\mathbf{k}_a,\mathbf{k}_b)
	=&  \int_{-L/2}^{L/2} \frac{\Xi}{L}
	\exp\left(-\tfrac{1}{8}w_0^2 |\mathbf{k}_a+\mathbf{k}_b|^2  \right. \nonumber \\
	&\left.-\frac{i z |\mathbf{k}_a-\mathbf{k}_b|^2}{2k_{\text{p}}}
	+i z\chi \right)\ \text{d}z . 
\end{align}
Next, we substitute $\mathbf{k}_a=-\mathbf{k}_b=\mathbf{k}_0$, so that
\begin{align}
	\eta(\mathbf{k}_0)
	= & \int_{-L/2}^{L/2} \frac{\Xi}{L}
	\exp\left(-\frac{i 2 z |\mathbf{k}_0|^2}{k_{\text{p}}}+i z\chi \right)\ \text{d}z .
	\label{eval2}
\end{align}
Finally, we evaluate the $z$-integration to get
\begin{equation}
	\eta(\mathbf{k}_0) = \Xi\text{sinc}\left(\frac{|\mathbf{k}_0|^2 L}{k_{\text{p}}}-\tfrac{1}{2} L\chi\right), 
	\label{eval3}
\end{equation}
which corresponds to the squeezing parameter when considering a realistic entanglement source.


\begin{thebibliography}{b}
	\bibitem{bennettel} C.H. Bennett, G. Brassard, C. Cr\'epeau, R. Jozsa, A. Peres, and W.K. Wootters, Teleporting an unknown quantum state via dual classical and Einstein-Podolsky-Rosen channels,
	{\em Phys. Rev. Lett}. {\bf 70}, 1895 (1993).
	\bibitem{bouwmeester1997experimental}D. Bouwmeester, J. Pan, K. Mattle, M. Eibl, H. Weinfurter, and A. Zeilinger,  Experimental quantum teleportation, {\em Nature}. \textbf{390}, 575-579 (1997).
	\bibitem{wang2015quantum} X. Wang, X. Cai, Z. Su, M. Chen, D. Wu, L. Li, N. Liu, C. Lu, and J. Pan, Quantum teleportation of multiple degrees of freedom of a single photon, {\em Nature}. \textbf{518}, 516-519 (2015).
	\bibitem{sephton2023quantum} B. Sephton, A. Vallés, I. Nape, M.A. Cox, F. Steinlechner, T. Konrad, F.S. Roux, and A. Forbes, Quantum transport of high-dimensional spatial information with a nonlinear detector, {\em Nature Communications} \textbf {14}, 8243 (2023).
	\bibitem{qiu2023remote} X. Qiu, H. Guo, and  L. Chen, Remote transport of high-dimensional orbital angular momentum states and ghost images via spatial-mode-engineered frequency conversion, {\em Nature Communications} \textbf {14}, 8244 (2023).
	\bibitem{vaidman1994} L. Vaidman,  Teleportation of quantum states, {\em  Phys. Rev. A} \textbf {49}, 1473 (1994).
	\bibitem{Braunstein0} S.L. Braunstein, and H.J. Kimble, Teleportation of continuous quantum variables, {\em Phys. Rev. Lett}. \textbf {80}, 869 (1998).
	\bibitem{lutkenhaus1999} N. Lütkenhaus, J. Calsamiglia, and K. Suominen,  Bell measurements for teleportation, {\em Physical Review A} \textbf{59}, 3295 (1999).
	\bibitem{goyal2014qudit} S. Goyal, P. Boukama-Dzoussi, S. Ghosh, F.S.  Roux, and T. Konrad, Qudit-teleportation for photons with linear optics, {\em Scientific Reports}. \textbf{4}, 4543 (2014).
	\bibitem{sokolov2001} I. Sokolov, M. Kolobov, A. Gatti, and L. Lugiato, Quantum holographic teleportation, {\em Optics Communications}. \textbf{193}, 175-180 (2001).
	\bibitem{Milburn} G.J. Milburn, and S.L. Braunstein, Quantum teleportation with squeezed vacuum states, {\em Phys. Rev. A} \textbf {60}, 937 (1999).
	\bibitem{Braunstein1} S.L. Braunstein, C.A. Fuchs, and H.J. Kimble, Criteria for continuous-variable quantum teleportation, {\em J.Mod.Opt}. \textbf{47}, 267 (2000).
	\bibitem{Hofmann} H.F. Hofmann, T. Ide, and T. Kobayashi, Fidelity and information in the quantum teleportation of continuous variables, {\em Phys. Rev. A} \textbf {62}, 062304 (2000).
	\bibitem{Braunstein2} S.L. Braunstein, and P. Van Loock, Quantum information with continuous variables, {\em Rev. Mod. Phys}. \textbf {77}, 513–577 (2005).
	\bibitem{Pirandola} S. Pirandola, and S. Mancini, Quantum teleportation with continuous variables: A survey, {\em Laser Physics} \textbf {16}, 1418 (2006).
	\bibitem{Asavanant} W. Asavanant, K. Takase, K. Fukui, M. Endo, J.-I. Yoshikawa, and A. Furusawa, Wave-function engineering via conditional quantum teleportation with a non-Gaussian entanglement resource, {\em Phys. Rev. A} \textbf {103}, 043701 (2021).
	\bibitem{Barnett} S. Barnett, and P. Radmore, {\it Methods in Theoretical Quantum Optics}, Oxford University Press, New York, (2003).
	\bibitem{Fabre} N. Fabre, Teleportation of continuous variables states, Master. Quantum Communication, Telecom ParisTech, France, 2023, pp.4. hal-03978754v2f.
	\bibitem{Miatto} F.M. Miatto, A.M. Yao, and S.M. Barnett, Full characterization of the quantum spiral bandwidth of entangled biphotons, {\em Phys. Rev.  A}  \textbf{83}, 033816 (2011).
	\bibitem{roux2020parametric} F.S. Roux, Parametric down-conversion beyond the semiclassical approximation, {\em Physical Review Research}. \textbf{2}, 033398 (2020).
\end{thebibliography}
\end{document}